\documentclass[twocolumn, nobibnotes, aps, 10pt, superscriptaddress, prb, longbibliography]{revtex4-1}
\usepackage{graphicx}
\usepackage{amsmath}
\usepackage{multirow}
\usepackage{longtable}
\usepackage{epstopdf}
\usepackage{color}
\bibpunct{[}{]}{,}{n}{}{}

\begin{document}

\title{Coulomb interactions and screening effects in few-layer black phosphorus: \\ a tight-binding consideration beyond the long-wavelength limit}

\author{D.~A. Prishchenko}
%
%\email[]{clasius@yandex.ru}
\affiliation{\mbox{Theoretical Physics and Applied Mathematics Department,
Ural Federal University, Mira Str. 19, 620002 Ekaterinburg, Russia}}

\author{V.~G. Mazurenko}
\affiliation{\mbox{Theoretical Physics and Applied Mathematics Department,
Ural Federal University, Mira Str. 19, 620002 Ekaterinburg, Russia}}

\author{M.~I. Katsnelson}
\affiliation{\mbox{Theoretical Physics and Applied Mathematics Department,
Ural Federal University, Mira Str. 19, 620002 Ekaterinburg, Russia}}
\affiliation{\mbox{Institute for Molecules and Materials, Radboud University, Heijendaalseweg 135, 6525 AJ Nijmegen, The Netherlands}}

\author{A.~N. Rudenko}
\email{a.rudenko@science.ru.nl}
\affiliation{\mbox{Theoretical Physics and Applied Mathematics Department, Ural Federal University, 
Mira Str. 19, 620002 Ekaterinburg, Russia}}
\affiliation{\mbox{Institute for Molecules and Materials, Radboud University, Heijendaalseweg 135, 6525 AJ Nijmegen, The Netherlands}}

\date{\today}

\begin{abstract}
Coulomb interaction and its screening play an important role in many physical phenomena of materials ranging from optical properties to many-body effects including superconductivity. Here, we report on a systematic study of dielectric screening in few-layer black phosphorus (BP), a two-dimensional material with promising electronic and optical characteristics. We use a combination of a tight-binding model and rigorously determined bare Coulomb interactions, which allows us to consider relevant microscopic effects beyond the long-wavelength limit. We calculate the dielectric function of few-layer BP in the random phase approximation and show that it exhibits strongly anisotropic behavior even in the static limit. We also estimate the strength of effective local and non-local Coulomb interactions and determine their doping dependence. We find that the $p_z$ states responsible for low-energy excitations in BP provide a moderate contribution to the screening, weakening the on-site Coulomb interaction by less that a factor of two.
Finally, we calculate the full plasmon spectrum of few-layer BP and discuss the effects beyond long-wavelengths.
\end{abstract}

%\pacs{}

\maketitle

\section{Introduction}

A few-layer black phosphorus (BP) has recently attracted significant attention as a prospective material for electronic and optical applications \cite{Ling,Peide,Gomez,Carvalho}. Two-dimensional (2D) BP features a direct layer-dependent energy gap, strongly anisotropic transport characteristic, and high carrier mobility \cite{LLi,Koenig,LHan,Qiao,Xia,Zant,rudenko_quasiparticle_2014, Tran,Rudenko_mobility}. These properties appear to be particularly promising for plasmonics, where BP is expected to show highly confined, low-loss, and tunable plasmon polaritons \cite{low_plasmons_2014,locplasm,Huber,edge}, as well as it offers the possibility to explore new plasmonic effects, such as hyperbolic optical response \cite{Nemilentsau,Serrano,LowNature}.

Electron-electron interactions and their screening are known to play a key role in determining the properties of materials. Those include optical properties ranging from the fundamental gap to collective excitations \cite{Onida}, charge carrier transport \cite{Sarma}, as well as more exotic examples such as superconductivity \cite{McMillan,Rietschel} and $sp$-magnetism \cite{Neto,Ziletti,Mazurenko}. Some aspects of the Coulomb screening in BP have already been addressed in the literature. Low \emph{et al.} \cite{low_plasmons_2014} studied screening in $n$-doped single-layer (1L) BP and despite highly anisotropic band dispersion found essentially isotropic screening in the static limit. In contrast, dynamical screening was found to exhibit strong anisotropy, enabling the existence of plasmons with anisotropic dispersion. Such prediction has been confirmed by electron energy loss spectroscopy (EELS), yet in the context of bulk BP crystal \cite{schuster_anisotropic_2015}. Similar theoretical results have been reported by Jin \emph{et al.} \cite{jin_screening_2015} with special emphasis on the effects of disorder in 1L- and 2L-BP. Other aspects of electron screening in 2D BP related to strain engineering of plasmons and to the Coulomb drag have been respectively addressed in Refs.~\onlinecite{lam_plasmonics_2015} and \onlinecite{pouya}.

Up to now, screening in BP has been studied on the basis of the random phase approximation (RPA) for the dielectric function \cite{low_plasmons_2014,jin_screening_2015,lam_plasmonics_2015,pouya}. Moreover, simple form of the Coulomb interaction in reciprocal space ($V\sim 1/q)$ was always assumed, which corresponds to the long-wavelength limit ($q \rightarrow 0$). Within this approximation, dielectric function is being considered as a scalar $\varepsilon(q)=1-V(q)\Pi(q)$, meaning that microscopic effects relevant at short distances are neglected. Although RPA polarizability $\Pi(q)$ is treated as a matrix within the tight-binding (TB) consideration \cite{jin_screening_2015,lam_plasmonics_2015}, this approximation is apparently not sufficient to capture all microscopic effects in $\varepsilon(q)$. At the same time, screening effects beyond the long-wavelength limit represent a problem of its own significance. For instance, they can be relevant for scattering processes involving large momentum transfer and are relevant in the context of many-body Hamiltonians. In 2D plasmonics, microscopic effects are especially relevant for interlayer interactions, giving rise to the existence of multiple plasmon modes \cite{Sarma,Pfnur}. To date, local Coulomb interaction in BP has only been considered as a parameter \cite{Ziletti}.

In this contribution, we study screening in a few-layer BP considering effects beyond the long-wavelength limit. We calculate static microscopic dielectric function of black phosphorus for different number of layers ($n=$1--3), as well as determine both short-range and long-range screening of the Coulomb interaction between the relevant BP orbitals. To this end, we use a tight-binding formalism in conjunction with the bare Coulomb interaction determined from realistic charge density distribution in real space. Based on the calculated microscopic dielectric function resolved over the whole Brillouin zone, we restore the plasmon spectra for each system considered and discuss their features arising beyond the limit $q \rightarrow 0$.

The paper is organized as follows. In Sec.~II, we consider bare Coulomb interactions in few-layer BP calculated from first-principles. Sec.~III is devoted to the static dielectric function calculated at the level of the TB model. Results on the static screened Coulomb interactions are presented in Sec.~IV. In Sec.~V, we analyze the spectrum of plasmon excitations in BP. In Sec.~VI, the paper is concluded.

\section{Bare Coulomb interaction}
\begin{figure}
\includegraphics[width=\columnwidth]{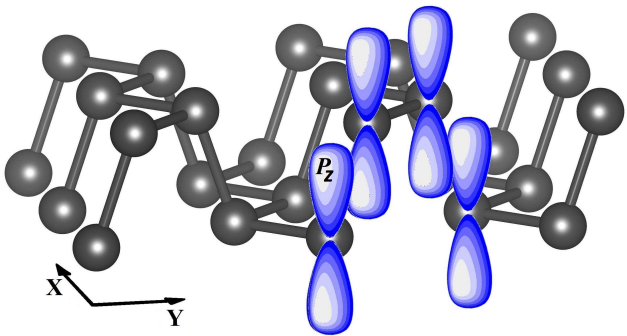}
\caption{%(Color online) 
Crystal structure of 1L-BP with schematic representation of $p_z$ orbitals.}
\label{struct}
\end{figure}

The classical form of the 2D Coulomb interaction, $V(q)=2\pi e^2/q$, is inapplicable for wave vectors $q\gtrsim 1/a_0$, where $a_0$ is the unit cell size. To adequately describe the interactions beyond the long-wavelength limit, both radial and angular dependence of the electronic density distribution at short distances should be taken into account.  Given that the relevant low-energy states in black phosphorus are predominantly composed of $p_z$ orbitals (schematically shown in Fig.~\ref{struct}), the bare Coulomb interaction between single-occupied orbitals residing at sites $i$ and $j$ can be calculated as
\begin{equation}
V_{ij} = e^2 \int d{\bf r} d{\bf r'} |w_i({\bf r})|^2 |{\bf r}-{\bf r}'|^{-1} |w_j({\bf r'})|^2,
\label{bareV}
\end{equation}
where $w_i(\bf r)$ is the Wannier function obtained by projecting the Bloch functions on the orbitals of $p_z$ symmetry localized on phosphorus atoms. Here, the Wannier functions are obtained within the formalism of projected Wannier functions \cite{Marzari} on the basis of $GW_0$ calculations performed in Ref.~\onlinecite{rudenko_toward_2015}. To this end, we use {\sc vasp} package \cite{vasp1,vasp2,vasp3} in conjunction with {\sc wannier90} code \cite{wannier90}. Integrals appearing in Eq.~(\ref{bareV}) are evaluated numerically.

\begin{figure}
\includegraphics[width=\columnwidth]{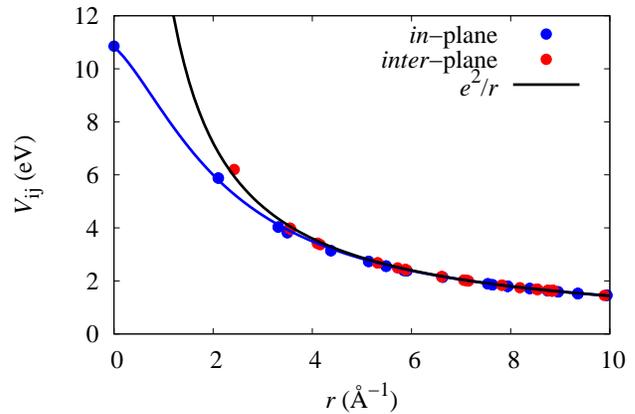}
\caption{\label{pcoul}%(Color online) 
Bare Coulomb interaction $V_{ij}(r)$ calculated between the $p_{z}$-like orbitals in BP. Blue points correspond to the interactions between the orbitals lying in-plane, while red points denote the interactions between the orbitals belonging to different planes. 
In-plane interactions are interpolated by blue line. Black line is the classical Coulomb law $e^2/r$.}
\end{figure}

Fig.~\ref{pcoul} shows the bare Coulomb interaction calculated between the $p_{z}$-like orbitals of phosphorus atoms in bulk BP for different distances. Keeping in mind strong angular dependence of $p_z$-like orbitals depicted schematically in Fig.~\ref{struct}, we distinguish between the in-plane and out-of-plane interactions. One can see that at short distances, calculated in-plane interaction (blue line) diverge from $2e^2/r$ (black line). To be able to use this data in our calculation of the dielectric function, we transform $V_{ij}$ to the reciprocal space,
\begin{equation}
V_{ij}(\textbf{q})=\sum_{\bf R} V_{ij}(\textbf{R})e^{-i\textbf{q}\textbf{R}} + \frac{2\pi e^2}{|{\bf q}|} - \sum_{\bf R}\frac{e^2}{|{\bf r}_{ij}+{\bf R}|}e^{-i\textbf{q}\textbf{R}},
\label{Vbareq}
\end{equation}
where $i$ and $j$ label now the atoms within the unit cell, and the sum runs over a finite cluster of unit cells separated by distance $|{\bf R}|$. In Eq.~(\ref{Vbareq}), the first term describes the short-range part of the interaction, whereas the second and third term ensure numerically accurate behavior of $V_{ij}({\bf q})$ in the long-wavelength limit (${\bf q}\rightarrow 0$).

\section{Dielectric function}

Within the random-phase approximation (RPA), frequency-dependent dielectric matrix $\epsilon_{ij}(\textbf{q},\omega)$ defined in terms of the momentum transfer vector ${\bf q}$ can be written as \cite{graf_electromagnetic_1995}:
\begin{equation}
\epsilon_{ij}(\textbf{q},\omega)=\delta_{ij}-\sum_{k} V_{ik}({\bf q})\Pi_{kj}(\textbf{q},\omega),
\label{epsilon}
\end{equation}
where $V_{ik}({\bf q})$ is the matrix of Coulomb interactions defined above in Sec.~II~A, and $\Pi_{kj}(\textbf{q},\omega)$ is the single-partice polarizability matrix. We note that in the present consideration, local field effects are neglected. 
In the basis of localized orbitals or Wannier functions, the polarizability matrix reads \cite{graf_electromagnetic_1995}:
\begin{multline}
\begin{split}
\Pi_{ij}&(\textbf{q},\omega)=\frac{g_{s}}{\Omega_0}\sum\limits_{{\bf k},mn}\frac{f_{m}(\textbf{k})-f_{n}(\textbf{k}+\textbf{q})}{E_{m}(\textbf{k})-E_{n}(\textbf{k}+\textbf{q})+\hbar \omega+i\eta} \\
& \times C_{i,m}(\textbf{k})C^{*}_{i,n}(\textbf{k}+\textbf{q})C^{*}_{j,m}(\textbf{k})C_{j,n}(\textbf{k}+\textbf{q}),
\end{split}
\label{Pi}
\end{multline}
where $C_{i,m}(\textbf{k})$ is the contribution of the $i$-th Wannier function $w_{i{\bf R}}({\bf r})$ to the Hamiltonian eigenstate $\psi_{m{\bf k}}({\bf r})$ with energy $E_{m}({\bf k})$:
\begin{equation}
\psi_{m{\bf k}}({\bf r})=\sum_i C_{i,m}({\bf k}) \phi_{i{\bf k}}({\bf r}),
\label{basis}
\end{equation}
where 
\begin{equation}
\phi_{i{\bf k}}({\bf r})=\sum_{\bf R}e^{i{\bf k}\cdot {\bf R}}w_{i{\bf R}}({\bf r}).
\end{equation}

In Eq.~(\ref{Pi}), $g_{s}=2$ is the spin degeneracy factor, $\Omega_0$ is the unit cell volume, $f_m({\bf k})=(\mathrm{exp}[(E_m({\bf k})-\mu)/k_{B}T]+1)^{-1}$ is Fermi-Dirac occupation factor, $\mu$ is the chemical potential determined by the carrier concentration $n$, and $\eta$ is a broadening term. In our calculations, we used $T=300$ K, $\eta=5$ meV, and $n=10^{13}$ cm$^{-2}$ both for electron and hole doping unless stated otherwise. Brillouin zone integration has been performed on a grid of $\sim$10$^6$ {\bf k}-points.

\begin{figure}
\includegraphics[width=\columnwidth]{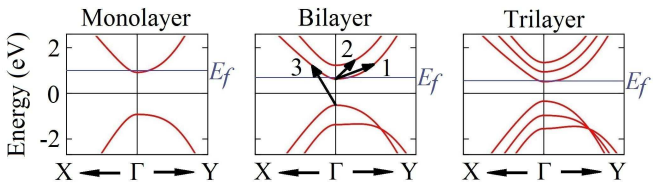}
\caption{\label{bandstr}% (Color online) 
Band structure of 1L-, 2L-, and 3L-BP obtained using the TB model used in this work. Arrows schematically show (1) intra-band, (2) inter-band, and (3) electron-hole pair excitations. Electron doping is shown as an example.}
\label{band2}
\end{figure}

\begin{figure*}
\includegraphics[width=2.0\columnwidth]{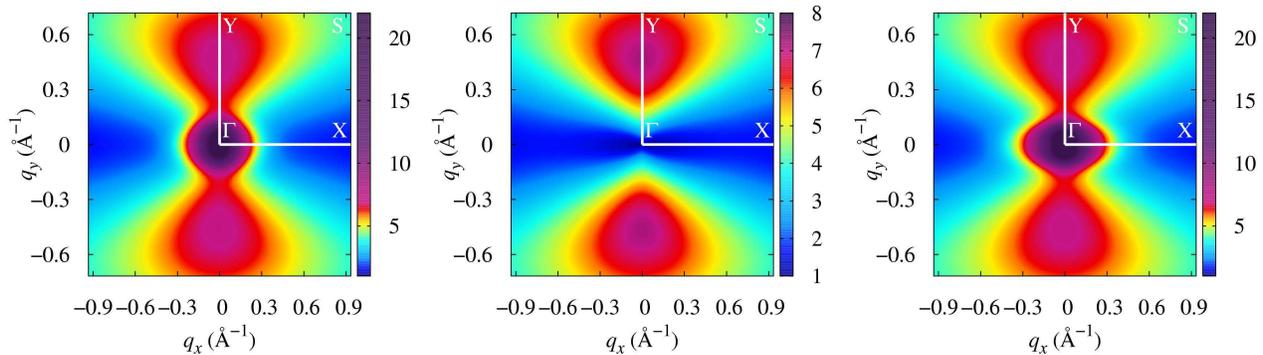}
\caption{%(Color online) 
Macroscopic static dielectric function $\epsilon_{M}(q_x,q_y)$ calculated for electron-doped (left), undoped (middle) and hole-doped (right) 1L-BP. Each plot shows distribution of $\epsilon_{M}$ over the whole BZ. Doping in both electron- and hole-doped cases corresponds to $n=10^{13}$ cm$^{-2}$.}
\label{diel}
\end{figure*}

Numerical determination of the polarizablilty matrix $\Pi_{ij}(\textbf{q},\omega)$ requires 
the knowledge of eigenvalues $E_m({\bf k})$ and eigenstates $\psi_{m{\bf k}}({\bf r})$ of a single-particle Hamiltonian. Those must be defined on a dense {\bf k}-point grid, ensuring accurate evaluation of the Brollouin zone integrals. Such calculations cannot be easily done directly from first principles. 
The method of model Hamiltonians offers an efficient alternative for studying electronic properties in BP. 
In our work, we use a tight-binding model proposed for multilayer BP in Ref.~\onlinecite{rudenko_toward_2015}, which is given by an effective Hamiltonian,
describing one electron per lattice site,
\begin{equation}
H=\sum_{i\neq j} t^{||}_{ij} c_i^{\dag}c_j + \sum_{i \neq j} t^{\perp}_{ij}c_i^{\dag}c_j,
\label{tb_hamilt}
\end{equation}
where $i$ and $j$ run over the lattice sites, $t^{||}_{ij}$ ($t^{\perp}_{ij}$) is intralayer (interlayer)  hopping integral between the $i$ and $j$ sites, and $c_i^{\dag}$ ($c_j$) is the
creation (annihilation) operator of electrons at site $i$ ($j$). The hopping integrals $t_{ij}$ are parametrized in Ref.~\onlinecite{rudenko_toward_2015} on the basis of first-principles $GW_0$ calculations, so that the model accurately describes the quasiparticle bands in the vicinity of the gap formed predominantly by the states of $p_z$ symmetry. The model is valid in a wide (up to visible light) spectral region and it is applicable to BP with arbitrary number of layers.  The model band structure and possible excitations in the vicinity of a band gap are shown in Fig.~\ref{band2} for 1L-, 2L-, and 3L-BP.

The macroscopic dielectric function $\epsilon_M({\bf q},\omega)$ can be calculated from the full dielectric matrix [Eq.~(\ref{epsilon})] by averaging over all the orbital states \cite{graf_electromagnetic_1995},
\begin{equation}
\epsilon_M(\textbf{q},\omega)=\frac{1}{N}\sum_{\textit{ij}}\epsilon_{ij}(\textbf{q},\omega),
\end{equation}
where $N$ is the number of states per unit cell.

Fig.~\ref{diel} shows static macroscopic dielectric function $\epsilon_{M}({\bf q}) \equiv \epsilon_{M}({\bf q},0)$ calculated for undoped, as well as $n$- and $p$-doped monolayer BP. Overall, one can see that $\epsilon_{M}$ is highly anisotropic, which primarily determined by the anisotropy of the polarizability matrix $\Pi_{ij}({\bf q},0)$. At the edges of the BZ, the screening anisotropy reaches its maximum, yielding
$\epsilon_{M}(Y)/\epsilon_{M}(X) \approx 4$ in all cases. 
The screening anisotropy is therefore more pronounced for wave vectors comparable with inverse unit cell size, $|{\bf q}|\sim 1/a_0$.
On the contrary, in the vicinity of the zone center ($\Gamma$ point) the dielectric function is essentially isotropic with $\epsilon_M({\bf q}\rightarrow 0)\rightarrow \infty$ for doped BP and $\epsilon_M({\bf q}\rightarrow 0)= 1$ in the absence of doping, which is consistent with generic dielectric properties of 2D materials \cite{Katsnelson-Book}. Indeed, for the doped case the result for small $q$ corresponds to the Thomas-Fermi approximation yielding $\varepsilon({\bf q})=1+\kappa/|{\bf q}|$, where $\kappa=2\pi e^2 N(\varepsilon_F)$ is the screening wave vector and $N(\varepsilon_F)$ is the density of states. Without doping, the polarization function can be represented at small $q$ in the general form as $\Pi({\bf q})=\sum_{\alpha \beta} C_{\alpha\beta} q_{\alpha} q_{\beta}$, where $C_{\alpha \beta}$ is some finite tensor. Noting that $V\sim 1/q$, one can see from Eq.~(\ref{epsilon}) that $\varepsilon_{M}({\bf q}\rightarrow 0)\rightarrow 1$.
We note that the behavior of the macroscopic dielectric function over the whole BZ does not change qualitatively with increasing the number of layers in BP.

\section{Screened Coulomb interaction}

With the knowledge of the bare Coulomb interaction  $V_{ij}(\bf q)$ and the dielectric matrix $\epsilon_{ij}({\bf q},\omega)$, the screened Coulomb interaction matrix $W_{ij}(\textbf{q})$ can be routinely calculated in the static limit ($\omega=0$) as
\begin{equation}
W_{ij}(\textbf{q})=\sum\limits_{p}\epsilon^{-1}_{ip}(\textbf{q},0)V_{pj}({\bf q}).
\label{W}
\end{equation}
$W_{ij}({\bf q})$ should be understood as 
\begin{equation}
W_{ij}({\bf q})=N_{k}^{-2}\sum_{{\bf k}{\bf k}'}\langle \phi_{i{\bf k}}({\bf r})\phi_{j{\bf k}'}({\bf r})| W | \phi_{i{\bf k}+{\bf q}}({\bf r}')\phi_{j{\bf k}'-{\bf q}}({\bf r}') \rangle,
\label{W2}
\end{equation}
where $N_k$ is the number of {\bf k}-points in the Brillouin zone, and
\begin{equation}
\langle ...| W |... \rangle = \! \int \! d{\bf r} d{\bf r}' \phi^*_{i{\bf k}}({\bf r}) \phi_{i{\bf k}+{\bf q}}({\bf r}) W  \phi^*_{j{\bf k}'}({\bf r}') \phi_{j{\bf k}'-{\bf q}}({\bf r}')
\end{equation}
is the matrix element of the screened static interaction $W=W({\bf r},{\bf r}';0)$,
describing the interaction of electrons with orbital indices $i,j$ and momenta ${\bf k}$,${\bf k}'$, which involves momentum transfer ${\bf q}$.

\begin{figure}
\includegraphics[width=\columnwidth]{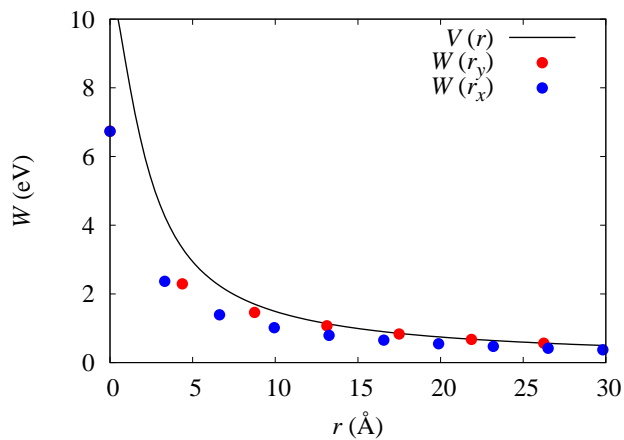}
\caption{\label{coulsc1}% (Color online) 
Diagonal element of the screened Coulomb interaction matrix $W$ calculated in real space along $x-$ (blue) and $y-$ (red) directions of 1L-BP. Unscreened (bare) interaction $V(r)$ is shown for comparison.}
\end{figure}

\begin{figure}
\includegraphics[width=\columnwidth]{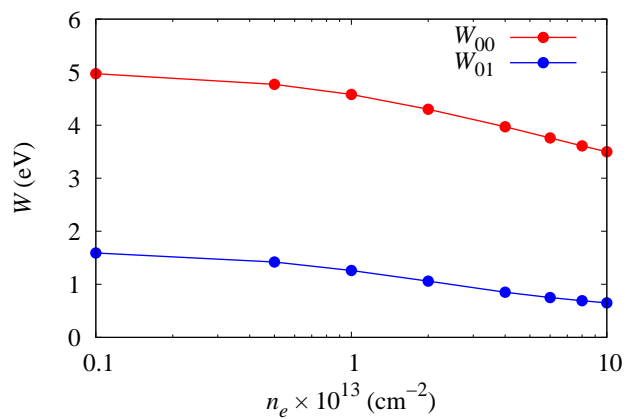}
\caption{%(Color online) 
On-site ($W_{00}$) and nearest-neighbor ($W_{01}$) screened Coulomb interaction in 1L-BP shown as a function of electron doping $n_e$. Red and blue lines are guide for the eye.}
\label{w-dop}
\end{figure}

\begin{table}
\caption{\label{coultab1} Computed on-site and nearest-neighbor bare and screened Coulomb interactions in BP within the same layer.}
\begin{ruledtabular}
\begin{tabular}{cccccc}
\multirow{2}{*}{$W_{ij}$(eV)} & \multirow{2}{*}{\textit{Bare}} & \multicolumn{3}{c}{\textit{Screened}} & \multirow{2}{*}{$R_{ij}$(\r{A})} \\
 &  & \textit{1 layer} & \textit{2 layers} & \textit{3 layers} & \\
 \hline
$W_{00}$  & 10.85 & 6.74 & 5.96 & 5.64 & 0.0 \\
$W_{01}$  &  5.88 & 3.27 & 2.73 & 2.63 & 2.22 \\
$W_{02}$  &  6.20 & 5.06 & 4.30 & 4.07 & 2.24 \\
$W_{03}$  &  4.03 & 2.37 & 1.88 & 1.84 & 3.31 \\
$W_{04}$  &  3.82 & 2.88 & 2.30 & 2.21 & 3.47 \\
\end{tabular}
\end{ruledtabular}
\end{table}

\begin{figure*}
\includegraphics[width=2\columnwidth]{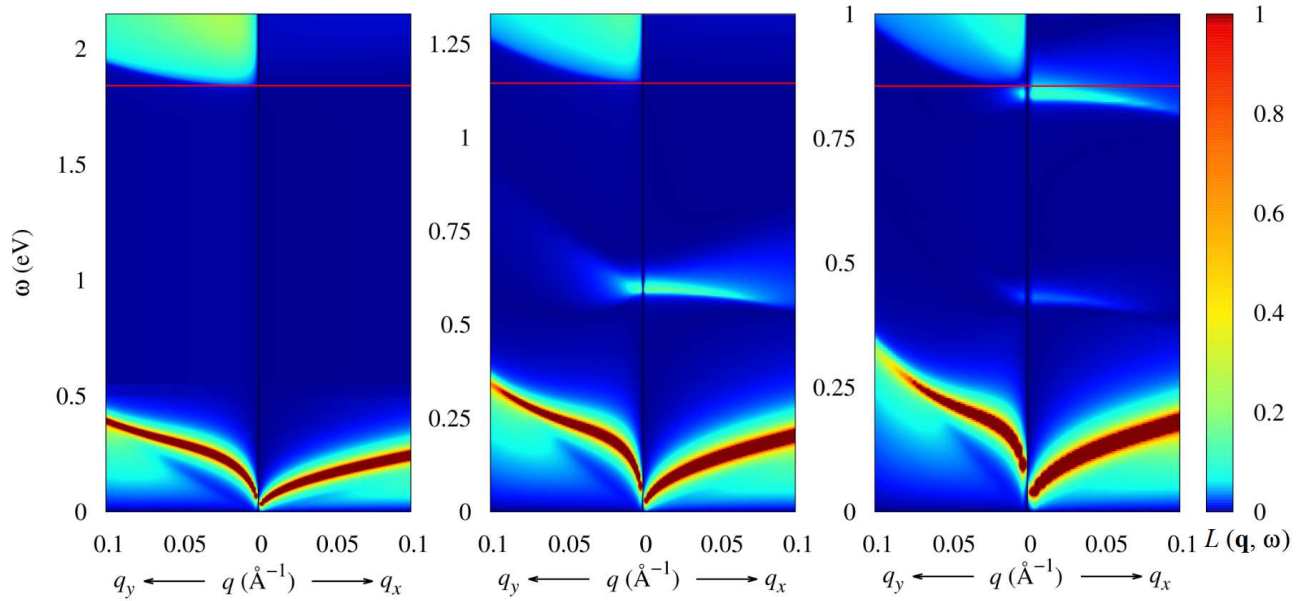}
\caption{\label{spectra}%(Color online) 
Wave vector and frequency resolved loss function $L(\textbf{q},\omega)$ (shown in color) calculated for 1L-, 2L-, and 3L-BP. Left and right part of each spectrum corresponds to $y-$ and $x-$ direction, respectively. Red horizontal line marks an energy gap for each system. Note different scales along vertical ($\omega$) axis. In all cases, electron-doping was used, corresponding to $n_{e}=10^{13}$ cm$^{-2}$.}
\end{figure*}

Fig.~\ref{coulsc1} shows calculated screened Coulomb interaction $W_{00}$ in undoped 1L-BP. The screening is more efficient at short distances ($W_{00}/V_{00}\approx$ 0.6), while at distances $r \gtrsim 10$ \AA, the interaction is virtually unscreened. Screening anisotropy is less pronounced compared to the reciprocal space (Fig.~\ref{diel}), making the interactions along the zigzag ($x$) direction slightly smaller than those along the armchair ($y$) direction. On-site ($W_{00}$) and nearest-neighbor $W_{01}$ screened interaction in 1L-BP is estimated to be 6.7 and 3.3 eV, respectively, which is somewhat larger than the fully screened interaction predicted for graphene \cite{Wehling}. We note, however, that here we are focused on the screening effects originating exclusively from the $p_z$ states of phosphorus, whereas other states of $p$ symmetry as well as high energy states are neglected. In Fig.~\ref{w-dop}, the dependence of $W_{00}$ and $W_{01}$ on the electron doping $n_e$ in 1L-BP is shown. One can see that doping enhances the screening significantly. At experimentally achievable gate doping of $n_e=10^{14}$ cm$^{-2}$, $W_{00}$ and $W_{01}$ reach 3.5 and 0.7 eV, respectively. Qualitatively the same results are obtained for the case of hole doping (not shown here).

Table~\ref{coultab1} shows calculated screened Coulomb interactions for BP with different (1--3) number of layers. As expected, the interaction strength decreases with the number of layers. This can be attributed to a larger screening associated with a reduced band gap in multilayer BP. Indeed, for smaller band gaps $\Delta$ interband transitions provide larger contributions to the polarization function as $\Pi_{ij}({\bf q},0)\sim 1/\Delta$ [Eq.~(\ref{Pi})].

\section{Plasmons}

\begin{figure*}
\includegraphics[width=2\columnwidth]{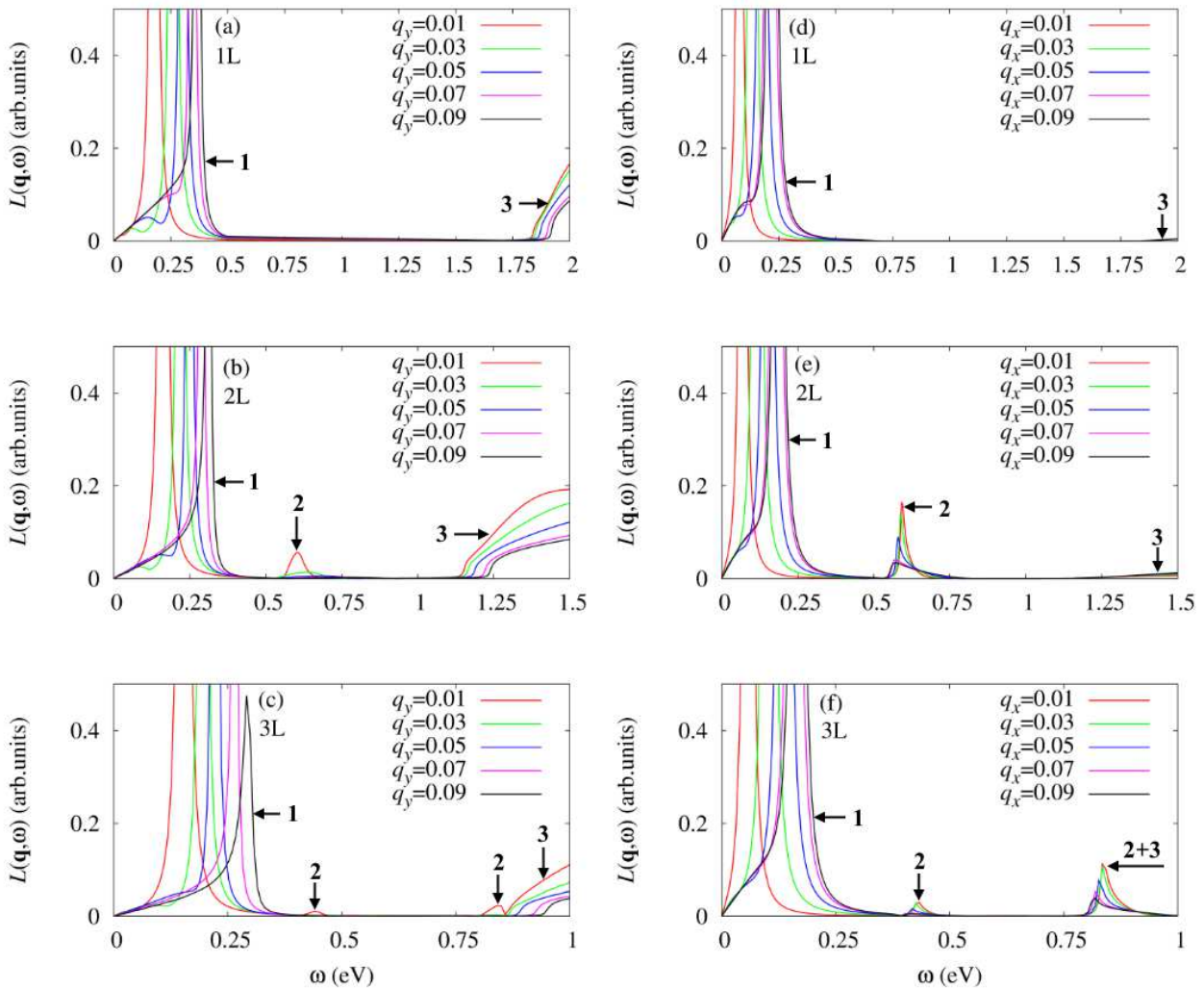}
\caption{\label{coulscr}% (Color online) 
Frequency-dependent loss function $L({\bf q},\omega)$ calculated for 1L-, 2L-, and 3L-BP for different wave vectors $q$ along $y-$ [(a)-(c)] and $x-$ [(d)-(f)] directions. Wave vectors are given in \AA$^{-1}$. Arrows show three different types of excitations in accordance with Fig.~\ref{band2}.}
\end{figure*}

In the absence of external field, the following equation serve as a criterion for the existance of self-sustained oscillations of the charge density (plasmons),
\begin{equation}
\int\epsilon(\omega,\textbf{r},\textbf{r}')\phi(\omega,\textbf{r}')d\textbf{r}'=0,
\label{plasmons1}
\end{equation}
where $\epsilon(\omega,\textbf{r},\textbf{r}')$ is the dielectric function in a continuous representation and $\phi(\omega,\textbf{r}')$ is the potential, induced by charge oscillations.
In the discrete representation, Eq.~(\ref{plasmons1}) can be rewritten in the form of a secular equation,
\begin{equation}
\mathrm{det}[\epsilon_{mn}({\bf q},\omega)]=0,
\label{plasmons2}
\end{equation}
which defines the plasmon spectrum of a system. In the presence of the plasmon damping, it is more convenient to obtain the plasmon spectrum by finding the maxima of the loss funcion $L({\bf q},\omega)=-\mathrm{Im}[1/\prod_{i}\epsilon_{i}(\omega)]$, with $\epsilon_{i}(\omega)$ being the $i$-th eigenvalue of $\epsilon_{mn}({\bf q},\omega)$.

Fig.~\ref{spectra} show the loss function $L({\bf q},\omega)$ of 1-, 2- and 3L-BP calculated for wave vectors up to 0.1 \AA$^{-1}$ resolved over the $x$ and $y$ directions, and for excitation energies $\omega>\Delta$, where $\Delta$ is the energy gap. For each number of layers considered, one can see the prominent $\omega \sim {\sqrt q}$ dependence, which is the classical plasmon dependence originating from the intraband transitions (type 1 in Fig.~\ref{bandstr}) in 2D electron gas. The corresponding dispersion is anisotropic, being suppressed in the y-direction as a consequence of the anisotropy of the BP electronic structure. Similar results have been previously obtained in the long-wavelength limit for 1L- \cite{low_plasmons_2014,lam_plasmonics_2015,jin_screening_2015} and 2L-BP \cite{jin_screening_2015,pouya} using low-energy continuum and TB Hamiltonians. 

Deviations from the $\omega \sim {\sqrt q}$ dependence occur at $q>0.05$ \AA$^{-1}$ and become more pronounced as the number of layer increases. Apart from the main plasmon mode, there are additional damped excitations at $\omega < {\sqrt q}$, whose intensity $L(q,\omega) \sim q$ in both crystallographic directions, as can be seen from Fig.~\ref{coulscr} for all the systems considered. This kind excitations emerge predominantly at $q>0.05$ \AA$^{-1}$ and thus are typical to short wavelengths. Technically, the origin of those modes is related to the presence of nondiagonal elements in the polarizability matrix $\Pi_{ij}({\bf q},\omega)$ [Eq.~(\ref{Pi})], and thus can be interpreted as out-of-phase oscillations of electronic density within or between the layers. Similar behavior has been predicted in metallic bilayers of transition metal dichalcogenides \cite{andersen_plasmons_2013}. The description of such features in BP requires to go beyond the continuum low-energy models and is possible at the TB level \cite{jin_screening_2015}.

In the long-wavelength limit ($q\rightarrow 0$), a charged layer induces a Coulomb potential $v(q)$ decaying with distance $z$ as $e^{-qz}$ \cite{ferrell}. In the context of 2D materials, this results in the emergence of an acoustic plasmon mode with dispersion $\omega \sim q$ in bilayer materials \cite{sensarma}. Such mode has been shown to exist in 2L-BP \cite{jin_screening_2015,pouya}, though it turns out to be strongly damped similar to bilayer graphene \cite{sensarma,tlow}. In our calculations, we do not make any assumptions regarding the behavior of the intralayer Coulomb potentials. Having rigorously calculated the Coulomb interaction matrix, $V_{ij}({\bf q})$ [Eq.~(\ref{Vbareq})], we observe strongly damped acoustic plasmon mode at small $q$ for all the cases under consideration \emph{including} 1L-BP. It is worth noting that the energies reported here for plasmon excitations associated with intraband transitions are somewhat underestimated. This is due to the underestimation of screening effects neglecting the transitions between the states not included into the model Hamiltonian. 

For multilayer BP, another type of excitations comes into play, namely, weakly dispersed optical plasmon mode appearing at $\omega \approx 0.6$ eV for 2L-BP, as well as at $\omega_1 \approx 0.4$ eV and $\omega_2 \approx 0.8$ eV for 3L-BP (see Figs.~\ref{spectra} and \ref{coulscr}). The corresponding frequencies are close to the intraband resonance resulting from the transitions between the subbands (type 2 on Fig.~\ref{bandstr}), whose splitting is governed by the interlayer interactions. Optical plasmon modes of comparable frequencies have been previously reported for bilayer graphene \cite{tlow}. It is interesting to note that the damping of optical plasmon excitations in BP is strongly anisotropic. While in $y$-direction the corresponding excitations decay at $q\sim 0.01$ \AA$^{-1}$, they are preserved up to $q \sim 0.06$ \AA$^{-1}$ in the $x$-direction. Although even at small $q$ optical plasmon is damped, its spectral weight is comparable with excitations in the particle-hole continuum ($\omega > \Delta$). Moreover, the spectral weight can be increased by increasing carrier doping or choosing a proper dielectric environment \cite{tlow}.

Optical excitations appearing in Figs.~\ref{spectra} and \ref{coulscr} at $\omega > \Delta$ are not directly related to the screening effects. They are observable without doping and originate from the dipole transitions. Strong anisotropy of this kind of excitations in 1L- and multilayer BP has been extensively analyzed previously using different theoretical methods \cite{tlow2,yuan2015,rudenko_toward_2015}, as well as observed experimentally in bulk BP \cite{schuster_anisotropic_2015}.

\section{Conclusion}
In this work, we studied dielectric screening and related properties of few-layer BP. We determined dielectric function matrix elements in the orbital subspace using the tight-binding model and a rigorous form of the bare Coulomb interaction computed in real space. Our consideration does not impose any restriction on the wave vector length and, therefore, goes beyond the long-wavelength limit studied before. 

Using the random phase approximation, we calculated the static dielectric function of few-layer BP over the whole Brillouin zone, which exhibits strongly anisotropic behavior, especially pronounced at the zone edges. The effective local and non-local Coulomb interactions screened by the $p_z$ orbitals included in the TB model are also estimated. Screening is shown to be more efficient at short distances, where the bare interaction is reduced by factor of two, and can be further increased by doping. In real-space, however, the anisotropy of the screened interaction is less evident.

Finally, we calculated the full plasmon spectrum for few-layer BP and classified the origin of different types of excitations. Short-wavelength effects related to the interactions between different sublattices are clearly observable in the plasmon spectrum of all the systems considered including 1L-BP. Apart from the classical $\omega \sim {\sqrt q}$ plasmon excitations, we observe additional quasi-linear ``acoustic'' plasmon mode originating from out-of-phase charge oscillations, as well as a weakly dispersed anisotropic ``optical'' resonance associated with interband transitions in multilayer BP.

The results presented here provide insights into the dielectric screening in BP at the microscopic level and can serve as a starting point for the analysis of many-body phenomena related to electron-electron coupling, such intrinsic charge carrier transport and superconductivity.
\newline
\begin{acknowledgments}
The work was supported by the grant program of the Russian Science Foundation 15-12-20021. Funding from the European Union's Horizon 2020 Programme under Grant No. 696656 Graphene Core1 is also gratefully acknowledged.
\end{acknowledgments}

\bibliography{bibl}
\end{document}